\documentclass[osajnl,twocolumn,showpacs,superscriptaddress,10pt]{revtex4-1} 
\usepackage{amsmath,amssymb,graphicx}
\begin{document}

\title{Integrated all-optical manipulation of orbital angular momentum carrying modes via enhanced electro-optic Kerr effect}

\author{S. Faezeh Mousavi}
\author{Rahman Nouroozi}\email{Corresponding author: rahman.nouroozi@iasbs.ac.ir}

\affiliation{Department of Physics, Institute for Advanced Studies in Basic Sciences (IASBS), Zanjan 45137-66731, Iran}

\begin{abstract} Mode Division Multiplexing (MDM) technique using higher order Orbital Angular Momentum (OAM) carrying modes through a channelized bandwidth provides enhanced capacity communication systems. OAM based high-dimensional Quantum Key Distribution (QKD) encrypted channels also improve transmission rate and security. All-optical mode-selective spatial distribution manipulation is a significant function in implemented MDM and QKD networks. This paper proposes a novel versatile-designed integrated optical device with Y$_{cut}$ ridge Periodically Poled Lithium Niobate (PPLN) photonic wire configuration which acts as spatial mode converter for data modulated on higher order OAM$=\pm2\hbar$ modes. It is schemed in such a way that control the phase of decomposed guided modes by enhanced electro-optic Kerr effect via phase-mismatched cascaded polarization coupling interaction in PPLN sections. The low-loss, high-purity (91 \%), and low-voltage proposed device enables to operate compatible with applicable commercial modulators in OAM based MDM and QKD communication systems.
\end{abstract}


\maketitle 

\section{Introduction}
Wavelength Division Multiplexing (WDM) and Polarization Division Multiplexing (PDM) are efficient techniques which occupy the fiber bandwidth by independent wavelength and polarization modulated channels, respectively, in order to meet the rapidly growing demands of high transmission capacities in optical networks \cite{PDMWDM,PDMWDM2,PDMWDM3}. However, the combination of Shannon limit and fiber nonlinearities prohibits larger capacity by these technologies \cite{mitra2001nonlinear}. Hence, Space Division Multiplexing (SDM) which utilizes the optimized cross section of the fiber with distinct channels in multi-core and multi-mode outlines, is a potential candidate of breakthrough technology against the capacity crunch \cite{9}. More specifically, Mode Division Multiplexing (MDM) in multi-mode fibers is an effective method with more achievable performance which exploits orthogonal spatial modes as new degrees of freedom for separately data modulation in independent channels of the fiber bandwidth \cite{39}. One of the recently interested high order spatial modes in communication systems is Orbital Angular Momentum (OAM) with helical phase structure of $e^{il\phi}$ ($\phi$, the azimuthal angle and $l=0,\pm1, \pm2, \ldots$ topological charge) which carries OAM of $l\hbar$ per optical mode \cite{yao2011orbital}. As a theoretically unbounded quantity of $l$, it has been proved that the combination of different OAM spatial modes can considerably expand the limits of traffic capacity in free space and fiber based optical networks \cite{40,10,14}. 

Meanwhile, in response to the security concern of communication systems, Quantum Cryptography (QC) improves privacy, authentication, and confidentiality for users. Quantum Key Distribution (QKD) protocols are effective approaches for commercialization of QC task. QKD schemes conventionally use a qubit system for distributing encoded information between two authorized participants connected by the quantum channel \cite{gisin2002quantum,scarani2009security}. Beyond two-level bases, it has been shown that employing multi-level quantum states can increase the robustness of a QKD system against eavesdropping \cite{bechmann2000quantum,cerf2002security}. High-dimensional QKD is feasible by OAM modulated states and provides higher transmission rates and security \cite{42,mirhosseini2015high}.

Accordingly, higher order OAM modes play key roles for realization of enhanced capacity and security via MDM and QKD systems in classical and quantum regimes \cite{42,41}. Therefore, as future prospects for the next generations of improved telecommunication and encrypted systems, mode-selective manipulation (such as (de)multiplexing, sorting, frequency conversion, polarization conversion, switch, and ...) of modulated information by OAM carriers is an inevitable function in actualized systems.

Practically, high-speed all-optically manipulation of encoded data in conventional communication optical channels can be precisely achieved by electro-optical and/or nonlinear effects in integrated optical devices. For instance, wavelength, phase, amplitude, and polarization of the electric field for a communication signal can be all-optically modulated in a miniaturized integrated waveguides \cite{lu2001wide,Chang2009}. Hitherto, several on-chip approaches such as asymmetric Y-junction \cite{27}, micro-ring \cite{MDMWDM_ADC} fiber based \cite{28,fusedfiber}, multimode interference \cite{30,WDMMDM_MMI}, adiabatic \cite{31}, multistaged \cite{29}, tapered directional \cite{TDC}, grating-assisted \cite{grating},  and asymmetrical directional \cite{32} couplings have been proposed to convert and multiplex the spatial modes. More specifically, some integrated optical modulators have been theoretically and experimentally investigated to emit, (de)multiplexe, switch, receive, detect, and sort OAM carrying modes \cite{emitter1,emitter2,emitter3,emitter4,emitter5,emitter6,multiplexing2,multiplexing3,multiplexing4,receiver,detector,sorter}. Even more, an integrated optical modulator has been proposed to alter the polarization and rotation handedness of $|l|=\pm1$ OAM carrying modes via manipulation of their decomposed guided modes by linear electro-optic effect \cite{24}. However, although higher ($|1|>1$) order OAM carrying modes impressively enable to provide even more enhanced capacity and improved security communication systems, but their integrated optically manipulation is not operated yet, to the best of our knowledge.  

\begin{figure}[htbp]
	\centering
	\includegraphics[width=1\linewidth]{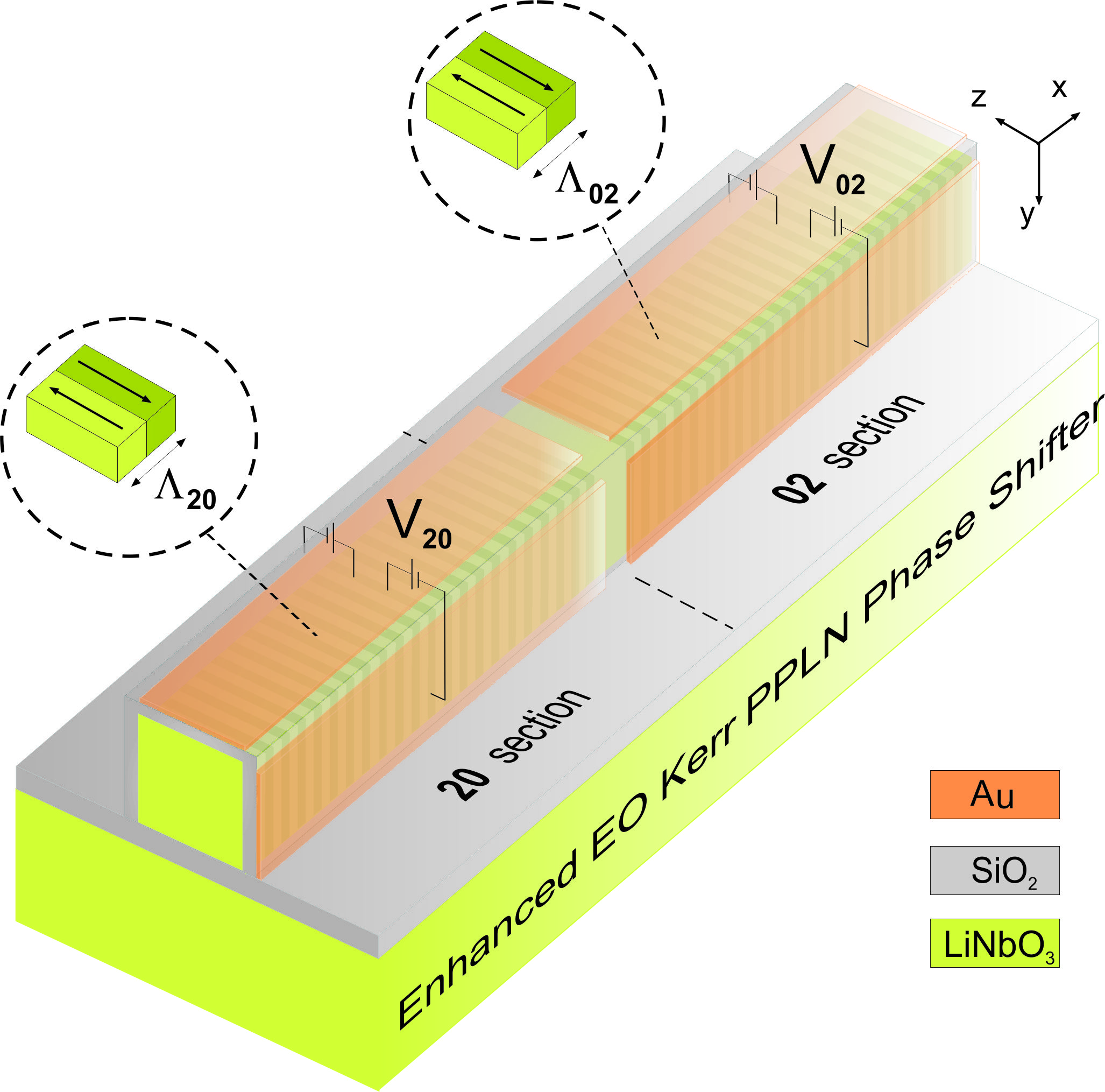}
	\caption{Schematic representation of the proposed device for spatial exchange of OAM$=\pm2\hbar$ modes. Green: LN (1.4 $\mu m$$\times$1.4 $\mu m$), gray: silica (200 $nm$ (top, lateral), 500 $nm$ (bottom)), and orange: gold electrodes (50 $nm$). 20 and 02 sections electro-optically modulate the interacting orthogonally polarized modes via enhanced electro-optic Kerr effect in phase-mismatched polarization coupling interactions to optimally induce the desired relative phases between three input $TE_{20,02,11}$ decomposed guided modes.}
	\label{fig:1}
\end{figure}

This paper proposes a versatile-designed integrated optical configuration enabling all-optically exchange of the spatial OAM modes with $l=\pm2$ via electro-optically manipulation of their decomposed guided modes. Similar to the introduced basic device of reference \cite{24}, the principal outline of the improved proposed modulator schematically displayed in figure \ref{fig:1} is based on a Y$_{cut}$ Lithium Niobate (LN) on Insulator (silica) (LNOI) ridge photonic wire configuration, whereas LN and silica act as the core and cladding, respectively. The top and lateral gold electrodes are additionally coated for applying external electric field and benefiting nonlinear electro-optic effect. 
Laguerre Gaussian ($LG$) modes as famous OAM carriers are combined of Hermite Gaussian ($HG$) modes with desired amplitudes and relative phases. More specifically, horizontally ($TE$) (vertically ($TM$)) polarized $LG$ modes are the desired phase related compounds of similarly $TE$ ($TM$) polarized $HG_{01}$ and $HG_{10}$ modes for $l=\pm 1$ (i. e. $0.5[HG_{01}\pm i HG_{10}]$); and $HG_{20}$, $HG_{02}$, and $HG_{11}$ modes for $l=\pm 2$  (i. e. $0.5[(HG_{20}-HG_{02})\pm  i\sqrt{2} HG_{11}]$) \cite{25}. Although irradiated $TE$ ($TM$) polarized LG modes are not individually guided ones of the symmetric rectangular shaped designed device, but are decomposed into its $TE_{02}$ ($TM_{02}$), $TE_{20}$ ($TM_{20}$), and $TE_{11}$ ($TM_{11}$) guided modes which are comparable to the mentioned horizontally (vertically) polarized $HG_{20}$, $HG_{02}$, and $HG_{11}$ ones. Since the decomposed guided modes are simultaneously excited in the designed modulator, their separately manipulation and relative phase compensation can achieve desirably modulated modes which carry $l=\pm 2$ OAM with high purity. Herein, the manipulation of decomposed guided modes is suggested, for the first time, to be obtained via enhanced electro-optic Kerr effect in cascaded phase-mismatched polarization coupling interactions (in 20 and 02 PPLN sections of figure \ref{fig:1}). Whereas the expressed effect can be utilized over diverse wavelength and polarization states, the novel proposed device enables to compatibly operate in practical fiber-based and free-space MDM and QKD communication systems and at the following of any desired polarization and wavelength dependent modulators (such as wavelength and polarization converters) which employ higher order OAM$=\pm2\hbar$ modes.

\section{Spatial Mode Conversion}

Investigating the manipulation of higher order spatial $|l|=2$ OAM carrying modes, a typical TE polarized OAM mode with $l=2$ is considered to be converted into similarly polarized one but with $l=-2$. The mentioned modes consist of desired phase related three decomposed $TE_{20}$, $TE_{02}$ and $TE_{11}$ guided modes. Their relative phases at the beginning of the waveguide are in such a way that lead to $TE_{l= 2}=\frac{1}{2}[(TE_{20}-TE_{02})+ i\sqrt{2} TE_{11}]$ \cite{25}. Indeed, this kind of complex higher order spatial mode conversion ($TE_{l=2}\Leftrightarrow TE_{l= -2}$) can be achieved if the relative phases between three decomposed guided modes are properly accomplished as:
\begin{equation}
\begin{split}
&	\frac{1}{2}[(TE_{20}-TE_{02})+ i\sqrt{2} TE_{11}] \Leftrightarrow\\
& \frac{1}{2}[(TE_{20}-TE_{02})- i\sqrt{2} TE_{11}].
\end{split}
\label{eq:1}
\end{equation}
According to ref \cite{24}, such a phase compensation is electro-optically attainable. The proposed modulator of ref \cite{24} was designed for manipulation of (e. g. TE polarized) OAM$=\pm1\hbar$ modes which are decomposed into two guided ones and so included one phase shifter to induce the desired relative phase between $TE_{01}$ and $TE_{10}$ modes. Since at least two relative phases associated to $|l|=2$ OAM modes have to be satisfied simultaneously, the proposed scheme in ref \cite{24} is not proportional. The improved modulator in this paper, thus, consists of two phase shifters to satisfy two required relative phases between three decomposed guided modes which are adequate for generation of desired OAM$=\pm 2 \hbar$ modes. Furthermore, albeit spatial mode conversion of $|l|=1$ OAM modes can be carried out via linear electro-optic process which varies by first order of external applied electric field (and then voltage) \cite{24}, it is not sufficient to satisfy more than one relative phases conditions. Consequently, the essential relative phases of the input OAM carrying modes with $|l|=2$ topological charges are proposed, for the first time, to be fulfilled by strong nonlinear phase shifters based on enhanced electro-optic Kerr effect \cite{34}. 

The enhanced Kerr effect as a recently introduced effective electro-optic phenomenon can be realized in LN (as a second order nonlinear medium) via cascaded phase-mismatched polarization coupling interaction \cite{35}. The polarization coupling process is possible in Periodically Poled Lithium Niobate (PPLN) by perturbation in the dielectric constant tensor of LN due to the transversely applied external electric field \cite{47}. Based on the cascaded phase-mismatched polarization coupling, the effective Kerr constant which is similar to the electro-optical Kerr effect and determined by the transversely applied electric field, is several orders of magnitude larger than that in the classic counterparts, and hence leads to large nonlinear phase shift for the input wave \cite{35,huo2014active}. 

In the case of input $TE_{l=2}$ OAM carrying mode incidence into the polarization coupling interaction, the amplitude of decomposed $TE_{20}$, $TE_{02}$, and $TE_{11}$ guided modes are related to the orthogonally polarized $TM_{20}$, $TM_{02}$, and $TM_{11}$ ones via coupled-mode equations as:
\begin{equation}
\begin{cases}
\dfrac{dA_{1,j}}{dx}=-i{\kappa_{j}} {A_{2,j}}{e}^{i\Delta\beta_{j} x_{j}},\\
\\
\dfrac{dA_{2,j}}{dx}=-i{\kappa_{j}}^{*} {A_{1,j}}{e}^{-i\Delta\beta_{j} x_{j}}.
\end{cases}
\label{eq:2}
\end{equation}

In this equation, $A_{2,j}$ and $A_{1,j}$ are the amplitude of input TE ($TE_{j=20,02,11}$) and output TM ($TM_{j=20,02,11}$) decomposed guided modes which are determined by $\alpha_{eff_{j}}$ effective loss coefficients and refractive indices of $n_{eff,TE_{j}}$ and $n_{eff,TM_{j}}$, respectively.  $\Delta\beta_{j}=\frac{2\pi}{\lambda}(n_{eff,TM_{j}}-n_{eff,TE_{j}})-2\pi/\Lambda_{j}$ also represents the wave-vector mismatch and so phase-mismatch between two interacting orthogonal modes in $\lambda$ wavelength during $x_{j}$ long sections with $\Lambda_{j}$ poling wavelength of PPLN, and	 $\kappa_{j}={{2{n}^{2}_{eff,TM_{j}}{n}^{2}_{eff,TE_{j}}r_{51}\langle{E_{y}}\rangle \vartheta_{j}}/{\lambda {(n_{eff,TE_{j}} n_{eff,TM_{j}})}^{1/2}}}$ is the coupling coefficient. In the definition of $\kappa$, $r_{51}$ is the desired electro-optical coefficient of LN, and $\vartheta_{j}$ describes the overlap integral between spatial distribution of two interacting orthogonal optical modes and the external electric field applying in y direction ($E_{y}$) \cite{47,24}. 

In the limit of weak cascading and trivial depletion of input mode (under non quasi-phase matching (NQPM) condition), the total phase of $j^{th}$ input TE mode ($\varphi_{tot,TE_{j}}$) which consists of propagation $\varphi_{prop,TE_{j}}$ and electro-optic phases ($\varphi_{eo,TE_{j}}$), is modified dominantly by second power of applied external electric field ($E_{y}$) as:
\begin{equation}
\begin{split}
&	\varphi_{tot,TE_{j}}=\varphi_{prop,TE_{j}}+\varphi_{eo,TE_{j}}=\\
&	\frac{2\pi}{\lambda}[n_{TE_{j}}+\frac{1}{2} ({n_{TE_{j}}^{3}}r_{51} \langle{E_{y}}\rangle tan \theta_{j}+s_{eff} \langle{E_{y}}\rangle^{2})] x_{j}.
\end{split}
\label{eq:3}
\end{equation}

In this equation, $\theta_{j}=r_{51}\langle{E_{y}}\rangle/(1/n_{TE,j}^{2}-1/n_{TM,j}^{2})$ is the rotation angle of new index ellipsoid via applied external electric field and $s_{eff}=-2(n_{TE,j}n_{TM,j})^{3} r_{51}^{2}/\pi \lambda \Delta \beta_{j}$ represents the effective electro-optical Kerr constant of input mode \cite{34}. In order to satisfy the appropriate relative phase between three decomposed $TE_{j=20,02,11}$ guided modes which independently accumulate their own phases during the propagation through the device, only two independent PPLN sections with known lengths are thusly considered to weakly convert the orthogonal modes (according to the coupled equations \ref{eq:2}) and benefit large nonlinear phase in enhanced Kerr effect (according to equation \ref{eq:3}). These two sequent sections are arbitrarily selected and functionalized to negligibly perform ${TE_{20}}\Leftrightarrow{TM_{20}}$ and ${TE_{02}}\Leftrightarrow{TE_{02}}$ couplings. Thus, three equation \ref{eq:3} for $\varphi_{tot,TE_{20}}$, $\varphi_{tot,TE_{02}}$, and $\varphi_{tot,TE_{11}}$ are numerically evaluated. Then two $\varphi_{tot,TE_{20}}-\varphi_{tot,TE_{02}}$ and $\varphi_{tot,TE_{20}}-\varphi_{tot,TE_{11}}$ relative phases equations with two $V_{20}$ and $V_{02}$ unknowns are calculated in such a way to provide suitable relative phases between three $TE_{j=20,02,11}$ decomposed modes and achieve ${TE_{l=2}}\Leftrightarrow{TE_{l=-2}}$ conversion. The proposed device which is schematically displayed in Fig. \ref{fig:1} includes two identically 15 mm long PPLN sections for 20 and 02 spatial modes with $\Lambda_{20}=9.56$ $\mu m$ and $\Lambda_{02}=13.33$ $\mu m$ poling wavelengths, receptively. Such domain grating wavelengths are chosen to attain high phase-mismatches and weakly ${TE_{20}}\Leftrightarrow{TM_{20}}$ and  ${TE_{02}}\Leftrightarrow{TE_{02}}$ conversions through the cascaded polarization coupling processes in 20 and 02 PPLN sections. The numerically calculated voltages which consequently change TE modes of $l=2$ into $l=-2$ are $V_{20}=14.58$ and $V_{02}=9.91$ volts. Table \ref{tab:1} indicates the effective refractive indices and loss coefficients of the involved modes at $\lambda=850$ nm wavelength for the mentioned interactions.    

\begin{table}[h!]
	\caption{Characteristics of the involved modes at $\lambda=850$ nm wavelength for $TE_{l=2}\Leftrightarrow TE_{l=-2}$ spatial mode conversion.}
	\begin{center}
		\begin{tabular}{ccc}
			\hline
			$mode$ &$n_{eff}$&$\alpha_{eff}$ (dB/cm)\\
			\hline 
			$TE_{02}$&1.995532&0.57900\\
			$TE_{20}$&1.987716&0.02224\\
			$TE_{11}$&2.027477&0.11191\\
			$TM_{02}$&2.058751&0.10033\\
			$TM_{20}$&2.076045&0.00218\\
			$TM_{11}$&2.103931&0.01282\\
			\hline 
		\end{tabular}
	\end{center}
\label{tab:1}
\end{table}

\begin{figure}[h!]
\centerline{\includegraphics[width=\columnwidth]{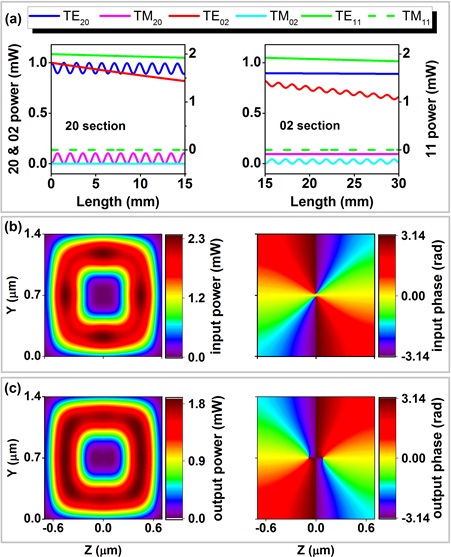}}
\caption{Calculated power evolution of the $20_{TE,TM}$ (blue), $02_{TE,TM}$ (red), and $11_{TE,TM}$ (green) involved guided modes in enhanced electro-optics Kerr effect via cascaded polarization coupling processes along the two 15 mm long 20 and 02 successive PPLN sections. Power distribution (left) and phase pattern (right) of input (b) and output (c) TE OAM mode with $l=2$ and $l=-2$ topological charges.}
\label{fig:2}
\end{figure}

Fig. \ref{fig:2}(a) illustrates the power evolution of the involved orthogonal guided modes along the 20 and 02 PPLN sections effected by the calculated voltages. As theoretically expected, in section 20 ${TE_{20}}\Leftrightarrow{TM_{20}}$ cascaded coupling is weakly attained by applied $V_{20}$ voltage while the other ${TE_{20}}$ and ${TE_{02}}$ input guided modes are not modified except attenuated. Followingly, in response to applied $V_{02}$ voltage, weakly ${TE_{02}}\Leftrightarrow{TM_{02}}$ coupling is achieved cascadingly in 02 PPLN section, whereas ${TE_{20}}$, ${TM_{20}}$ and ${TE_{11}}$ involved modes are wholly unchanged just by loss affects. Furthermore, the lengths of 20 and 02 PPLN sections are specified in a way to optimize adequate amplitude ratios of three ${TE_{20,02,11}}$ decomposed guided modes and generate output $|l|=2$ OAM carrying mode with efficient configuration. The power distribution (left) and phase pattern (right) of the input (OAM$=2\hbar$) and output (OAM$=-2\hbar$) modes are respectively displayed in Fig. \ref{fig:2}(b) and Fig. \ref{fig:2}(c). Their comparisons confirm that the generated output OAM mode carries $l=-2$ topological charge. This result is corroborated by 91 \% calculated purity of the output mode, too. They also point to the output mode production of 78 \% efficieny with respect to the expressed attenuations.

\section{Conclusion}
A novel integrated optical Y$_{cut}$ PPLN ridge photonic wire configuration is proposed and numerically evaluated to electro-optically manipulate the spatial higher order OAM$=\pm2\hbar$ modes. To meet the desired relative phases, the input decomposed guided modes are modified via enhanced electro-optic Kerr effect in cascaded phase-mismatched polarization coupling interaction through PPLN sections. The low-loss, low-voltage, and high-purity (91 \%) schemed device is compliant with free space and fiber based practical MDM and QKD communication systems which employ OAM modes in classic and quantum  regimes. It also enables to operate consecutively after many applicable operator for higher order OAM modes, such as polarization dependent mode-selective wavelength converter, (de)multiplexer, switch, and sorter.

\section{Supplemental Material}

For implementation of the proposed modulator in this paper, the procedures listed below are suggested \cite{gui2010periodically,hu2009lithium,rabiei2013heterogeneous,ueno2012entangled}:  

1) Exploiting plasma enhanced chemical vapor deposition (PECVD) method for coating silica layer on $Y_{cut}$ $LN$ substrate (first layer). This silica layer would be as the bottom clad of the phonic wire.

2) A $He^+$ ion implanted $Y_{cut}$ $LN$ layer (second layer) can be used to bond to the insulator on $LN$ layer (first layer) prepared in the first step.

3) Annealing the bonded layers to improve bonding strength and split thin layer of $Y_{cut}$ $LN$ along implanted ions. This layer that would be remained on the $SiO_{2}-LN$, is the core of modulator.

4) Applying Chemical Mechanical Polishing (CMP) on the films to improve the surface roughnesses of resulted $LN$ on Insulator (LNOI) wafer.

5) Using Plasma etching for slicing the $LN$ part of resulted LNOI wafer and achieving a core with rectangle cross section attached to bottom silica clad and $LN$ substrate.

6) Using photolithography for fabricating comb-like electrodes and then applying electric field to pole the etched $LN$ (PPLN) core periodically. For forming $PPLN$ in this modulator, structure with two part, any one with its specific wavelength is recommended. 

7) Exploiting PECVD method for coating thin layer of $SiO_2$, as top and lateral clads, on etched $LiNbO_3$ core; and then slicing the resulted coated wafer by plasma etching.

8) Applying PECVD method for coating thin layer of $Au$ on $SiO_2$  layer (top and lateral clads), as top and lateral electrodes.

9) Using Plasma etching to form $Au$ electrodes in the desired shapes.


%


\end{document}